\documentclass{PoS}

\usepackage[caption=false]{subfig}
\usepackage{graphicx}
\usepackage{bm}
\usepackage{amssymb}
\usepackage{amsmath}
\usepackage{amsfonts}
\usepackage[utf8]{inputenc}
\usepackage{placeins}

\usepackage{booktabs}
\usepackage{gensymb}

\usepackage{microtype}
\usepackage[utf8]{inputenc}

\usepackage{listings}
\usepackage{siunitx}
\usepackage{xspace}
\usepackage{comment}

\title{Searches for Connections Between Dark Matter and Neutrinos with the IceCube High-Energy Starting Event Sample}

\ShortTitle{HESE Dark Matter}

\author{
The IceCube Collaboration\footnote{For collaboration list, see PoS(ICRC2019) 1177.}\\
{\itshape \href{http://icecube.wisc.edu/collaboration/authors/icrc19_icecube}{http://icecube.wisc.edu/collaboration/authors/icrc19\_icecube}}\\
E-mail: \email{caad@mit.edu}
}

\abstract{
The nature of dark matter remains one of the most important open questions in physics. Although dark matter effects have only been observed gravitationally, the order-one ratio between conventional matter and dark matter hints to a non-gravitational link between them. In this contribution, we search for such a link using the IceCube high-energy starting event (HESE) sample, which contains some of the highest energy neutrinos ever observed. These are dominated by a yet unidentified high-energy diffuse astrophysical component. Using these events we look for evidence of an excess due to dark matter decay or annihilation, and also for a novel signature due to high-energy neutrinos scattering with dark matter. Finally, in this contribution, we will summarize and highlight the complementarity between the obtained limits for these two scenarios.
\vspace{4mm}
{\bfseries Corresponding authors:}
\speaker{Carlos A. Arg{\"u}elles}$^{1}$ and Hrvoje Dujmovic$^{2}$\\
{$^{1}$ \itshape Massachusetts Institute of Technology, Cambridge, MA 02139, USA}\\
{$^{2}$ \itshape Department of Physics, Sungkyunkwan University, Suwon, 16419, Korea}

}

\FullConference{36th International Cosmic Ray Conference -ICRC2019-\\
		July 24th - August 1st, 2019\\
		Madison, WI, U.S.A.}
		
\begin{document}

\section{Introduction}

The origin of neutrino masses and the nature of dark matter are the two most compelling evidences for physics beyond the Standard Model (SM). At the moment no model is as successful as the corpuscular dark matter hypothesis to explain the wide range of observational evidence for it from cosmology to astrophysical measurements. Alternative theories such as modified gravity are able to explain local measurements, but fail to match precision cosmological observables, {\it e.g.} PLANCK measurement of the cosmic microwave background~\cite{Ade:2015xua}, and are now significantly disfavored. Thus, in this work, we will assume that dark matter is a particle with unknown interactions with the SM particle content. The lightness of neutrino masses, approximately six orders of magnitude smaller than the electron, invites one to think of neutrino-mass-generation mechanisms that not only explain their non-zero value, but also their smallness~\cite{Mohapatra:2005wg}. In fact, the {\it scotogenic} -- from the greek {\it scoto} meaning darkness -- origin of neutrino mass has been proposed as a model for neutrino mass generation~\cite{Boehm:2006mi,Farzan:2012sa,Escudero:2016tzx,Escudero:2016ksa,Alvey:2019jzx}. More recently, heavy neutrinos -- also known as dark neutrinos depending on the context -- have been introduced to explain the MiniBooNE anomaly and also explain neutrino masses~\cite{Bertuzzo:2018itn,Ballett:2018ynz,Ballett:2019cqp,Ballett:2019pyw}. Within these models it is possible that neutrinos dominate the interaction between SM particles and dark matter; this is in opposition to the usual belief that due to gauge symmetry charged-lepton bounds dominate~\cite{Blennow:2019fhy}.

\section{Sample and analyses signatures\label{sec:ana}}

The results presented in this contribution use seven and a half years of the high-energy starting event (HESE) sample. This sample was introduced and is described in~\cite{Aartsen:2013bka,Aartsen:2014gkd}. The event selection uses the outer part of the detector as a veto region to reject atmospheric muons and neutrinos produced in showers initiated by high-energy cosmic-rays. Atmospheric muons with incident energy of 10~TeV  have approximately 10\% probability of passing the veto layer without triggering it, while muons at 100~TeV only have 1\% probability. These rejection probabilities, combined with the steeply falling atmospheric muon flux, yield an expected atmospheric muon rate of $\mathcal{O}(0.1)$ events above 60 TeV deposited energy for the analysis lifetime. Even though atmospheric neutrinos do not emit light as they cross the veto region, their expected contribution to the sample is suppressed due to the fact that muons produced in the same air shower have non-negligible chance of triggering the veto; for a recent discussion on this effect and its uncertainties see~\cite{Arguelles:2018awr}. For these reasons, the events above 60~TeV in this sample are expected to be dominated by extraterrestrial neutrinos; for an updated results of the astrophysical flux characterization using this sample see~\cite{yuan_tianlu_2018_1300506}.

\begin{figure*}
    \centering
    \includegraphics[width=0.45\linewidth]{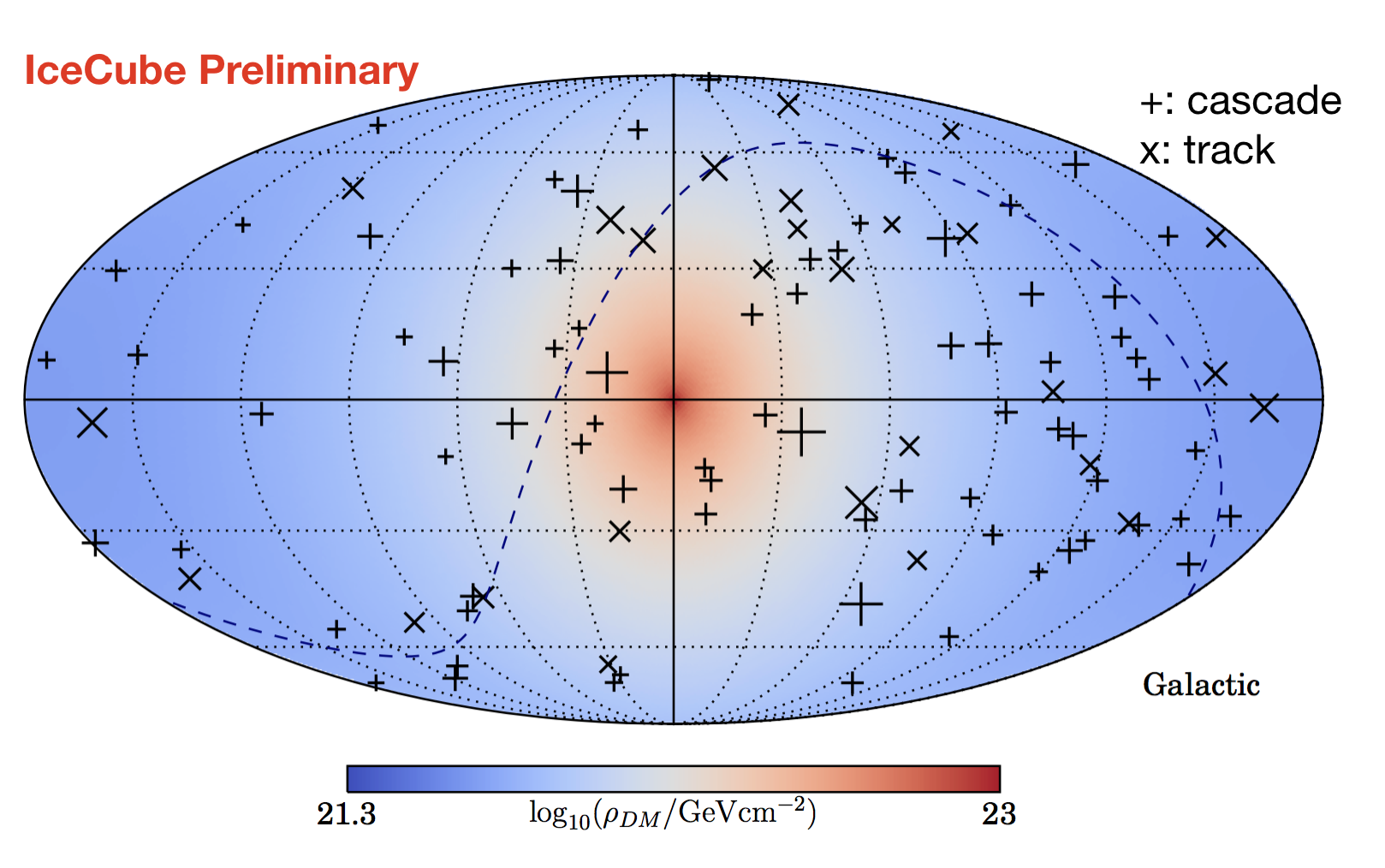}
    \caption{\textbf{\textit{Events on top of galactic dark matter distribution.}} High-energy starting events best-fit directions are shown in galactic coordinates. For track-like events the ``+'' marker is used, while for cascade-like events the ``x'' marker is used. The colored background is the galactic dark matter column density along that direction, increasing amount of dark matter density goes from blue to red. The marker sizes have no meaning.}
    \label{fig:neutrino-skymap}
\end{figure*}

Searches for signatures characteristic of dark matter in this sample can be organized in two categories. First, as evidence for an additional component of neutrinos observed on top of atmospheric and astrophysical backgrounds. This additional component can be produced from the dark matter decay or its annihilation into SM particles. These two scenarios differ in their expected angular and energy distributions. Since dark matter decay only depends linearly on the dark matter density, it exhibits a broader angular distribution; while dark matter annihilation depends quadratically on the dark matter density producing a more peaked distribution around the galactic center. Figure~\ref{fig:neutrino-skymap} shows the events in the sample, including events below 60~TeV, on top of the dark matter column density. The energy distribution of the neutrinos produced in the decay or annihilation varies significantly depending on the assumed decay or annihilation channel. We can group them into two broad categories depending on the neutrino energy spectrum: hard channels ($\nu\bar\nu$,$\mu^+\mu^-$,$\tau^+\tau^-$,$h\nu$)\footnote{$h$ refers to the Higgs boson.} and soft channels ($b\bar b$,$WW$). In complete theories, dark matter can decay (or annihilate) into one of these channels or a combination of them depending on the underlying theory.

The second category is as a modification of the incoming astrophysical neutrino flux direction, spectrum, or flavor~\cite{Arguelles:2019rbn,Farzan:2018pnk,Cherry:2014xra,Kelly:2018tyg,Arguelles:2017atb}. In particular, the astrophysical neutrino flux spatial distribution has been measured for signs of anisotropy. This search have return empty handed, since the data is compatible with an isotropic distribution, which is expected from far yet unresolved sources. In this case, as astrophysical neutrinos traverse the galaxy they can interact with the galactic dark matter. Due to the anisotropic distribution of dark matter in the sky, this imprints an anisotropy on the astrophysical neutrino flux observed at Earth~\cite{Arguelles:2017atb}. Since the interactions of dark matter and neutrinos are not known, interaction models of them need to be assumed. Due to the high neutrino beam energies and wide range of target dark matter masses an effective interaction description is not appropriate. Instead, we make use of two simplified models to study the possible interactions between neutrinos and dark matter. In one scenario we assume that dark matter is a fermion and can scatter off neutrinos via the exchange of a vector boson. In the other scenario, we assume that dark matter is a scalar and it can interact with neutrinos via the exchange of a fermionic mediator. The different neutrino-dark matter effective models and their cross sections can be found in~\cite{Campo:2017nwh}. Even though these simplified models are not manifestly gauge invariant, they can be embedded in complete theories in such a way that neutrino interactions are the dominant portal to dark matter~\cite{Blennow:2019fhy}. 

\section{Results}

\begin{figure*}[t]
\centering
\subfloat[Results on dark matter decay.]{
\includegraphics[width=0.50\linewidth]{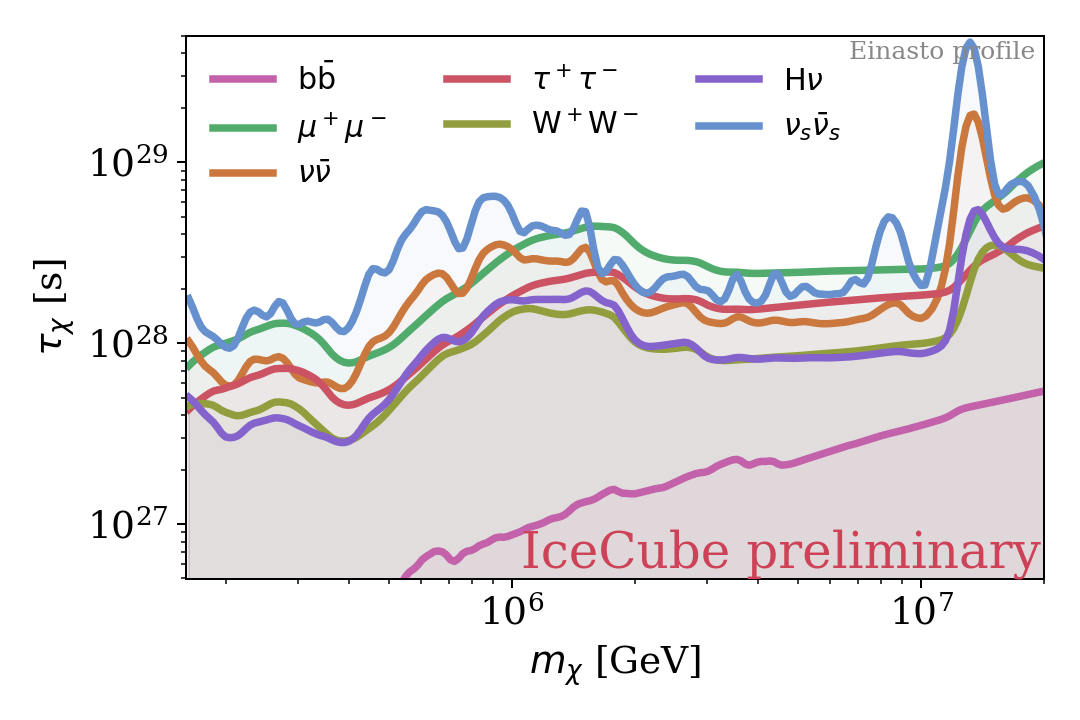}}~
\subfloat[Results on dark matter annihilation.]{
\includegraphics[width=0.50\linewidth]{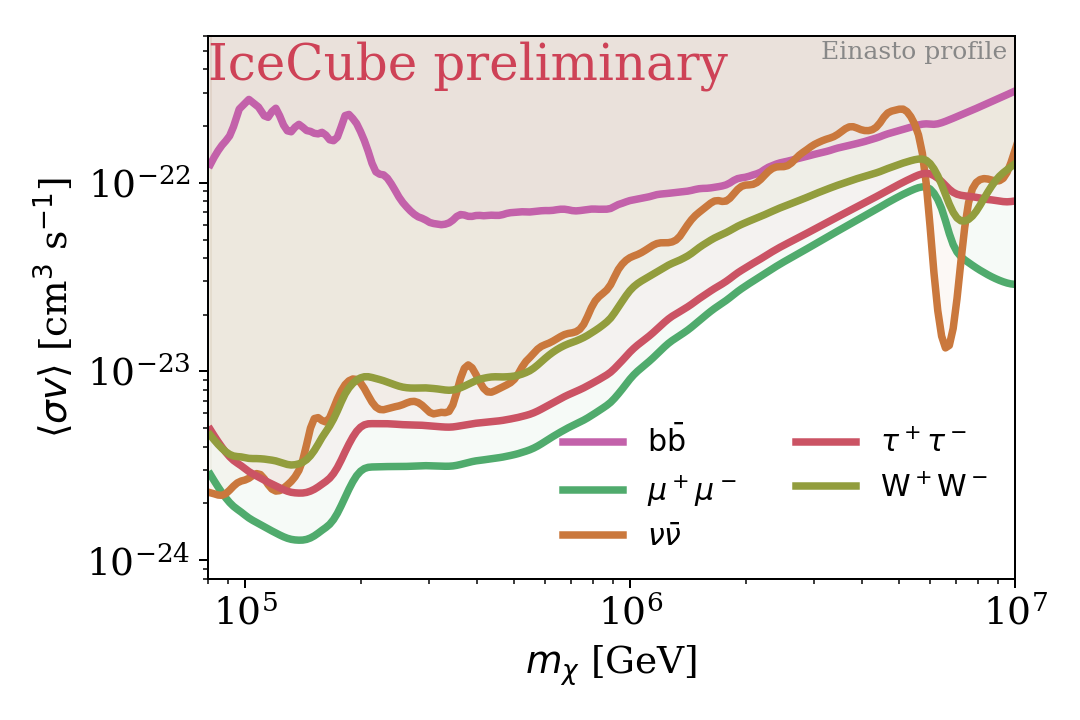}}\\
\subfloat[Results in dark matter scattering.]{
\includegraphics[width=0.50\linewidth]{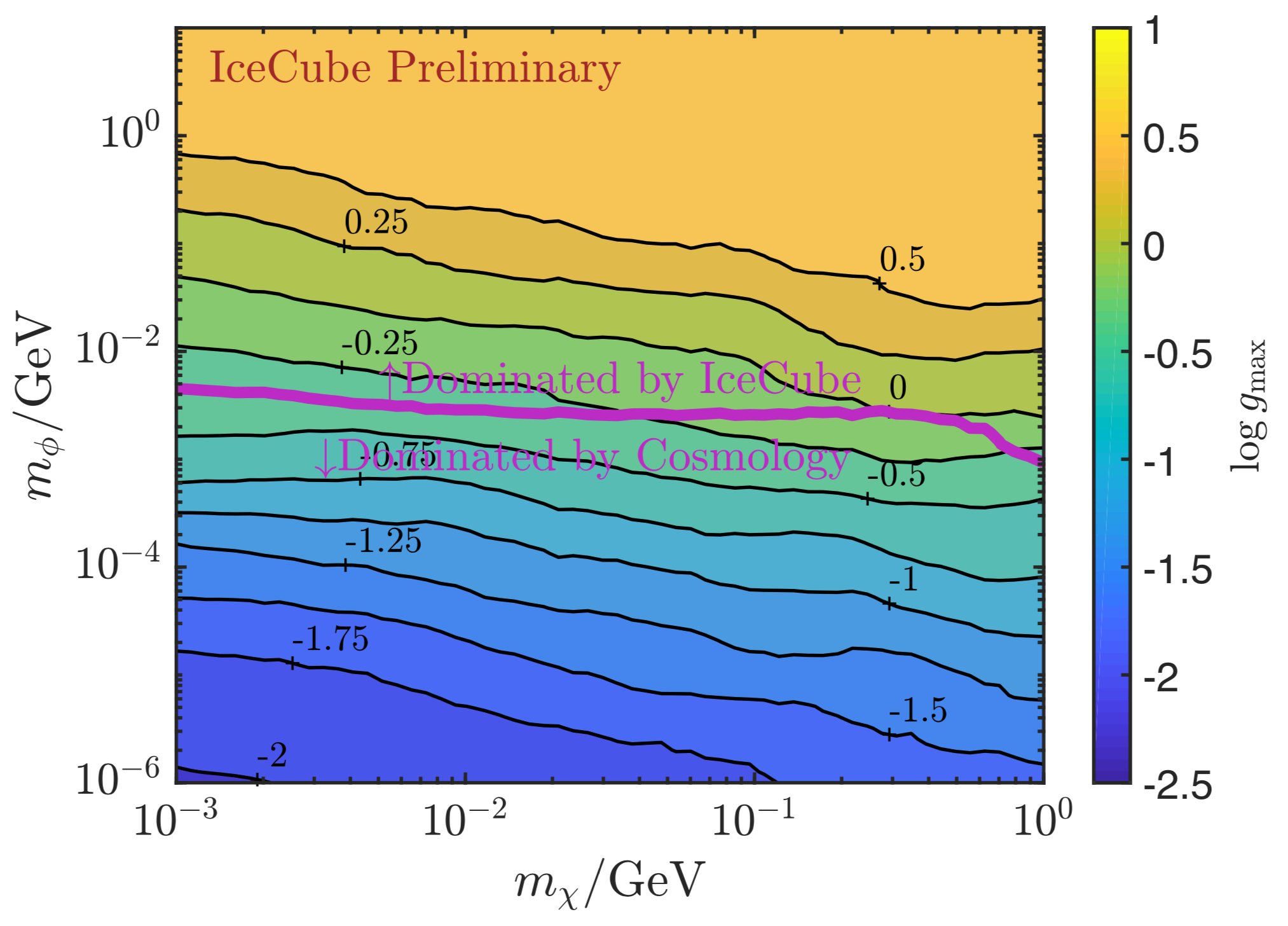}}
    \caption{\textbf{\textit{Constraints for dark matter decay, annihilation, and scattering with neutrinos.}} The three main results of this analysis are shown as a function of the dark matter mass, $m_\chi$. (a): Constraints on dark matter lifetime for different assumed decay modes shown as different line colors. (b): Constraints on the dark matter annihilation cross section to different standard model particles. (c): Constraints on the neutrino dark matter cross section via looking for signatures of neutrino scattering. In this analysis we assume the dark matter is a fermion and the interaction mediator, $\phi$, is a vector. The vertical axis shows the assumed mediator mass, $m_\phi$, and the color scale gives the maximum allowed log of the coupling. The magenta line signals the regions where cosmological observations are more constraining and where this analysis constraints are dominant.}\label{fig:results}
\end{figure*}

We performed the analyses outlined in Sec.~\ref{sec:ana} searching for evidence of dark matter decay, annihilation, or scattering with neutrinos and found no evidence for them. In these analyses, we bin data and simulation in right ascension and declination. We compute the expected rate due to atmospheric background, astrophysical flux, and a potential dark matter component in each bin. Data and simulation are then compared using the likelihood defined in~\cite{Arguelles:2019izp}, which accounts for the statistical uncertainty inherent in the simulation. In figure ~\ref{fig:results} we show the new constraints obtained. We discuss them here briefly:
\begin{itemize}
    \item Dark matter annihilation: The most stringest constraint are found for lepton annihilation channels -- $\nu\bar\nu$, $\mu^+\mu^-$, and $\tau^+\tau^-$. These are followed by the $W^+W^-$ channel, with the exception of energies where the $W^+$ decay products are close to the Glashow resonance. The $b\bar b$ channel is the weakest as it produces a soft spectrum closer to the observed astrophysical spectrum.
    \item Dark matter decay: The strongest constraints are found for hard channels such as channels $\nu_s\bar\nu_s$, introduced in~\cite{Brdar:2016thq}, $\nu\bar\nu$, and $\mu^+\mu^-$. An increased exclusion power is found when the dark matter mass is approximately twice the Glashow resonance~\cite{Glashow:1960zz}. The $b\bar b$ channel is the softest channel and similarly to~\cite{Aartsen:2018mxl} it produces the weakest limits. This is due to the fact that there is a partial degeneracy between a soft astrophysical component and this spectrum, resulting in a best-fit point that is preferred over null -- though not significant to claim evidence for dark matter.
    \item Dark matter-neutrino scattering: Results are shown for fermionic dark matter with a vector boson mediator. In this case the coupling that is being constrained is the product of the $\chi\chi\phi$ and $\nu\nu\phi$ vertices' couplings. The maximum 90\% credible upper limit of the square root of the product of these couplings is shown as a color log-scale as a function of the dark sector masses. IceCube's bounds are the leading bounds for mediator masses above 10~MeV for a broad range of dark matter masses from 1~MeV to 1~GeV; lower mediator masses are dominated by cosmological observations~\cite{Escudero:2015yka}.
\end{itemize}

\section{Conclusions}

We have performed a search for signals of neutrino-dark matter interactions. We have found no evidence for them. We have placed new constraints on dark matter annihilation, decay, and scattering. It is important to note that the constraints from scattering and annihilation are caused by diagrams related by crossing symmetries. Thus these analyses are complementary and put constraints in the dark matter-neutrino cross section from a few MeV to 10~PeV dark matter masses. A more detailed comparison can be made once a model is assumed, {\it e.g.} for the vector mediated dark matter-neutrino interaction the cross section is constrained to be less than $\mathcal{O}(10^{-1})$ barns for dark matter masses approximately less than 100~MeV and then its constraints to be approximately less than $\mathcal{O}(10^{-6})$ barns for dark matter masses above 10~GeV. The smaller mass region is limited by looking at modifications of the astrophysical neutrino flux, while the larger masses are bounded via searches for excess due to dark matter in IceCube; these include not only the ones presented in this work but also those in ~\cite{Aartsen:2017ulx} for instance.
\bibliographystyle{ICRC}
\bibliography{references}

\end{document}